\documentclass[aps,pra,showpacs,twocolumn,eqsecnum]{revtex4}
\begin{document}
\title{Quantum state filtering applied to the discrimination of Boolean
functions}
\author{J\'anos A. Bergou}
\author{Mark Hillery}
\affiliation{Department of Physics, Hunter College of the City 
University of New York, 695 Park Avenue, New York, NY 10021}

\date{\today}

\begin{abstract}
Quantum state filtering is a variant of the unambiguous state discrimination 
problem: the states are grouped in sets 
and we want to determine to which particular set a given input state belongs.
The simplest case, when the $N$ given states are divided into 
two subsets and the first set consists of one state only while the second 
consists of all of the remaining states, is termed quantum state 
filtering.  We derived previously the optimal strategy for the case of $N$ 
non-orthogonal states, $\{|\psi_{1} \rangle,\ldots , |\psi_{N} \rangle \}$, 
for distinguishing $|\psi_1 \rangle$ from the set $\{|\psi_2 \rangle,
\ldots,|\psi_N \rangle \}$ and the corresponding optimal success and failure 
probabilities.  In a previous paper [\prl {\bf 90}, 257901 (2003)], we 
sketched an appplication of the results to probabilistic quantum algorithms. 
Here we fill the gaps and give the complete derivation of the probabilstic 
quantum algorithm that can optimally distinguish between two classes of 
Boolean functions, that of the balanced functions and that of the biased 
functions. The algorithm is probabilistic, it fails sometimes but when it 
does it lets us know that it did. Our approach can be considered as a 
generalization of the Deutsch-Jozsa algorithm that was developed for the 
discrimination of balanced and constant Boolean functions.

\end{abstract}

\maketitle
\section{Introduction}
The Deutsch-Jozsa algorithm was one of the first quantum algorithms
\cite{deutsch}.  It makes it possible to perform on a quantum computer 
in one step a task that, on a classical computer, would require many 
steps.  In particular, one is given a ``black box'' that, when given an 
$n$ bit input, returns either $0$ or $1$.  The box simply evaluates a 
Boolean function, that is a function that maps $n$-bit binary numbers 
to the set $\{ 0,1\}$.  It is guaranteed that the functions are of one 
of two types, they are either constant, giving the same value for all 
inputs, or balanced, giving $0$ for half the inputs and $1$ for the 
other half.  The task is to determine whether the function we have 
been given is constant or balanced.  On a classical computer, we would, 
in the worst case, have to evaluate the function $2^{(n-1)}+1$ times to 
determine this.  On a quantum computer the determination can be made with 
certainty with only one evaluation of the function.

The quantum algorithm works by mapping the function onto a quantum state.
The quantum states for constant and balanced functions lie in orthogonal
subspaces, and they can, therefore, be distinguished perfectly.  An 
obvious question to ask is whether the method can be extended to distinguish
functions whose vectors do not lie in orthogonal subspaces.  Here we shall
show that this is indeed the case, and to do so we shall make use of a
procedure known as quantum state filtering \cite{bergou1,bergou2,bergou3}.  

Quantum state filtering considers the following problem.  We are given
a system that is in one of a known set of $N$ quantum states $\{\psi_{1},
\ldots \psi_{N}\}$.  Our task is to determine whether the system is in
$\psi_{1}$ or not, that is, whether it is in the set $\{\psi_{1}\}$ or
the set $\{\psi_{2}, \ldots \psi_{N}\}$.  This is what quantum state
filtering accomplishes.  It is a probabilistic procedure, it may fail,
but we know when it does.  The failure probability should be as small as
possible, and how this is accomplished depends on the \emph{a priori}
probabilities with which the states $\{\psi_{1}, \ldots \psi_{N}\}$
appear and the overlaps of $\psi_{1}$ with the other states.  In order to 
generalize the Deutsch-Jozsa algorithm, we can apply quantum state
filtering to the quantum states into which the Boolean functions are
mapped.

Quantum state filtering is an instance of mixed state discrimination, in
particular the discrimination of a pure state from a mixed state.  The
set of states $\{\psi_{2}, \ldots \psi_{N}\}$ along with their 
\emph{a priori} probabilities can be viewed as an ensemble, which
can be represented by a density matrix.  There are several different 
strategies one can adopt in discriminating mixed states.  One can
minimize the probability of making an error \cite{helstrom,holevo,yuen}.
A second strategy is that of unambiguous discrimination in which
one never makes a mistake in identifying the state, but one can sometimes
fail to obtain any information about the state one was given
\cite{rudolph}-\cite{raynal2}.  Quantum state filtering is an example
of this approach.  There are also hybrid strategies in which one can
make mistakes and one can also fail to obtain an answer 
\cite{fiurasek,eldar2}.  The addition of the option of failing to obtain
an answer makes it possible to obtain a smaller error probability than is
possible when the possibility of failing is not present.

The paper is organized as follows.  In the next section we briefly review 
quantum state filtering, following \cite{bergou3} but with a notation 
simplified for the present purposes.  In Section 3 we apply state filtering to
the problem of distinguishing balanced functions from particular
biased functions, i.e.\ functions that have a preponderance of zeroes 
or ones.  In Section 4 we provide a specific example of the POVM that
figures in the state filtering procedure, and we summarize our results
in the Conclusion. 

\section{Quantum state filtering}
As was mentioned in the Introduction, what we wish to do, is to 
determine whether a system we have been given is in the state $\psi_{1}$
or not, given that it must be in one of the states $\{\psi_{1},
\ldots \psi_{N}\}$ and that the state $\psi_{j}$ occurs with probability
$\eta_{j}$.  This can be accomplished in one of three ways, and
which way is the best depends on the \emph{a priori} probabilities and
the overlaps of the states.  Two of these methods are von Neumann
projective measurements, and the third is a POVM.  

The first projective measurement projects onto the subspace
orthogonal to $\psi_{1}$.  This is accomplished by the projection
operator
\begin{equation}
F^{(1)}=I-|\psi_{1}\rangle\langle\psi_{1}| .
\end{equation}
If we obtain the value $1$, then the procedure has succeeded, and we
know that the system we have was not in the state $\psi_{1}$.  If we
obtain $0$, then the procedure has failed.  This happens with a 
probability
\begin{equation}
Q_{1}=\eta_{1} + S  ,
\label{Qsqm1}
\end{equation}
where
\begin{equation}
S=\sum_{j=2}^{N}\eta_{j}|\langle\psi_{1}|\psi_{j}\rangle|^{2}
\end{equation}
is the average overlap between the two subsets. 

A second possibility is to split $|\psi_{1}\rangle$ into two
components, $|\psi_{1}\rangle = |\psi_{1}^{\perp}\rangle +
|\psi_{1}^{\parallel}\rangle$.  Here $|\psi_{1}^{\perp}\rangle$ is
orthogonal to the subspace, $\mathcal{H}_{2}$, that is spanned by the
vectors $|\psi_{2}\rangle, \ldots |\psi_{N}\rangle$, and
$|\psi_{1}^{\parallel}\rangle$ lies in $\mathcal{H}_{2}$. Their 
normalized versions are $|\tilde{\psi}_{1}^{\perp}\rangle =
|\psi_{1}^{\perp}\rangle / \|\psi_{1}^{\perp}\|$ and
$|\tilde{\psi}_{1}^{\parallel}\rangle = |\psi_{1}^{\parallel}\rangle /
\|\psi_{1}^{\parallel}\|$, respectively.  We now introduce the operator
\begin{equation}
F^{(2)}=|\tilde{\psi}_{1}^{\perp}\rangle\langle
\tilde{\psi}_{1}^{\perp}| -
(I - |\tilde{\psi}_{1}^{\perp}\rangle\langle\tilde{\psi}_{1}^{\perp}|
-|\tilde{\psi}_{1}^{\parallel}\rangle\langle
\tilde{\psi}_{1}^{\parallel}|),
\end{equation}
which has eigenvalues $1$, $0$, and $-1$.  If we measure $F^{(2)}$ and
obtain $1$, then 
the vector was $|\psi_{1}\rangle$, if we obtain $-1$, then the vector
was in the set $\{ |\psi_{2}\rangle , \dots |\psi_{N}\rangle\}$, and
if we obtain $0$, the procedure failed. In this case the probability
of failure, $Q_{2}$, is given by
\begin{equation}
\label{Qsqm2}
Q_{2}=\eta_{1}\|\psi_{1}^{\parallel}\|^{2} +
\frac{S}{\|\psi_{1}^{\parallel}\|^{2}}.
\end{equation}
Which of these two particular strategies is better is determined by
which of these two failure probabilities is smaller. In particular,
$Q_{1}>Q_{2}$ if $\eta_{1}\|\psi_{1}^{\parallel}\|^{2}
>S$, and vice versa.

The third measurement is based on a
positive-operator valued measure (POVM, \cite{kraus}), and it can do
better in an intermediate range of parameters.
The POVM can be implemented by a unitary evolution on
a larger space and a selective measurement. The larger space
consists of two orthogonal subspaces, the original system space and a
failure space. The idea is that the unitary evolution transforms
the input sets into orthogonal sets in the original system space
and maps them onto the same vector in the failure space.  A click in
the detector measuring along this vector
corresponds to failure of the procedure, since all inputs are
mapped onto the same output. A no-click corresponds to success
since now the non-orthogonal input sets are transformed into
orthogonal output sets in the system space.  The
one-dimensionality of the failure space follows from the
requirement that the filtering is optimum, as shown by
the following simple considerations. Suppose that
$\psi_{1}$ is mapped onto some vector in the failure space
and the inputs from the other set are mapped onto vectors that
have components perpendicular to this vector.  Then a single von
Neumann measurement along the orthogonal direction could identify
the input as being from the second set, i.e. further filtering
would be possible, lowering the failure probability and, contrary
to our assumption, our original filtering could not have been
optimal.

In particular, let ${\cal H}_{S}$ be the $D$-dimensional system
space spanned by the vectors $\{ |\psi_{1}\rangle ,\ldots
|\psi_{N}\rangle\}$ where, obviously, $D \leq N$.  We now embed
this space in a space of $D+1$ dimensions, ${\cal H}_{S+A}={\cal
H}_{S} \oplus {\cal H}_{A}$, where ${\cal H}_{A}$ is a
one-dimensional auxiliary Hilbert space, the failure space or ancilla.
The basis in this space is denoted by $|\phi_{A}\rangle$,
where $\|\phi_{A}\|=1$.  Thus, the
unitary evolution on ${\cal H}_{S+A}$ is specified by the
requirement that for any of the input states $|\psi_{j}\rangle$
($j=1,\ldots,N$) the final state has the structure
\begin{equation}
\label{unit}
|\psi_{j}\rangle_{out}=U|\psi_{j}\rangle = \sqrt{p_{j}}
 |\psi_{j}^{\prime}\rangle + \sqrt{q_{j}} e^{i\theta_{j}}
|\phi_{A}\rangle ,
\label{U}
\end{equation}
where $|\psi_{j}^{\prime}\rangle\in \mathcal{H}_{S}$, and
$\|\psi_{j}\|=1$
From unitarity  the relation, $p_{j}+q_{j}=1$, follows.
Furthermore, $p_{j}$ is the probability that the transformation
$|\psi_{j}\rangle \rightarrow |\psi^{\prime}_{j}\rangle$ succeeds
and $q_{j}$ is the probability that $|\psi_{j}\rangle$ is mapped
onto the state $|\phi_{A} \rangle$.  In order  to identify $p_{j}$
and $q_{j}$ with the state-specific success and failure
probability for quantum filtering we have to require that
\begin{equation}
\langle\psi^{\prime}_{1}| \psi^{\prime}_{j}\rangle = 0 ,
\label{ortho}
\end{equation}
for $j=2,\ldots N$.  We can now set up $D+1$ detectors in the
following way.  One of them is directed along
$|\psi_{1}^{\prime}\rangle$, $D-1$ along the remaining $D-1$
orthogonal directions in the original system space and the last one
along $|\phi_{A}\rangle$ in the failure space.  If any of the
system-space detectors clicks we can uniquely assign the state we were
given to one or the other subset and a click in the failure-space
detector indicates that filtering has failed.  It should be noted that, 
in general, for the unitary operator given in Eq.\ (\ref{U}) to exist 
we must have $D=N$ \cite{bergou2}.

In order to optimize the POVM, we have to determine those values of
$q_{j}$ in Eq. (\ref{U}) that yield the smallest average failure
probability $Q$, where
\begin{equation}
\label{Q}  
Q=\sum_{j=1}^{N}\eta_{j}q_{j} .
\end{equation}
Taking the scalar product of $U|\psi_{1}\rangle$
and $U|\psi_{j}\rangle$ in Eq. (\ref{U}), and using Eq.
(\ref{ortho}), gives
\begin{equation}
|\langle \psi_{1}|\psi_{j}\rangle|^{2}=q_{1} q_{j} ,
\label{qproduct}
\end{equation}
for $j=2,\ldots,N$, and Eq. (\ref{Q}) can be cast in the form
$Q(q_{1}) = \eta_{1} q_{1} +S/q_{1}$. Unitarity of the
transformation $U$ delivers the necessary condition that
$q_{1}$ must lie in the range $\|\psi_{1}^{\parallel}\|^{2} \leq
q_{1} \leq 1$. Details of the derivation, along with a discussion of
the sufficient conditions for the existence of $U$, can be found
in \cite{bergou2}. Provided that a POVM-solution
exists, the minimum of $Q(q_{1})$ is reached for $q_{1} =
\sqrt{S/\eta_{1}}$ and is given by
\begin{equation}
Q_{POVM} = 2 \sqrt{\eta_{1} S}.
\label{Qpovm}
\end{equation}

Thus, the failure probability for optimal unambiguous quantum
state filtering can be summarized as
\begin{eqnarray}
    \label{Qmin}
    Q = \left\{ \begin{array}{ll}
    2 \sqrt{\eta_{1} S}
    & \mbox{if
    $\eta_{1}\| \psi_{1}^{\parallel} \|^{4}
    \leq  S
    \leq \eta_{1}$} , \\
    \eta_{1}+S & \mbox{ if $S > \eta_{1}$} , \\
    \eta_{1}\|\psi_{1}^{\parallel}\|^{2}  + \frac{S}
    {\|\psi_{1}^{\parallel}\|^{2}} & \mbox{ if $S
    < \eta_{1} \|\psi_{1}^{\parallel} \|^{4}$} .
    \end{array}
    \right.
\end{eqnarray}
The first line represents the POVM result, Eq.\ (\ref{Qpovm}), and it
gives a smaller failure probability, in its range of validity, than the
von Neumann measurements, Eqs.\ (\ref{Qsqm1}) and Eq.\
(\ref{Qsqm2}). Outside of the POVM range
of validity we recover the von Neumann results. It should be noted
that for these results to hold, unlike for unambiguous state
discrimination, linear independence of all states is not
required. Instead, the less stringent requirement of the linear
independence of the sets is sufficient, in agreement with the findings
in \cite{zhang}.

It is useful to see what happens if the POVM is applied to a vector that
lies in the subspace, $\mathcal{H}_{2}$, spanned by the vectors 
$\{ \psi_{2}, \dots \psi_{N}\}$, but is not one of these vectors.  Let
\begin{equation}
|\psi\rangle = \sum_{j=2}^{N}c_{j}|\psi_{j}\rangle , 
\end{equation}
so that
\begin{equation}
U|\psi\rangle = \sqrt{p}|\psi^{\prime}\rangle +\sqrt{q}e^{i\theta}
|\phi_{A}\rangle ,
\end{equation}
where $\|\psi^{\prime}\| =1$, and
\begin{eqnarray}
\sqrt{p}|\psi^{\prime}\rangle & = & \sum_{j=2}^{N}c_{j}\sqrt{p_{j}}
|\psi^{\prime}_{j}\rangle  \nonumber \\
\sqrt{q}e^{i\theta} & = & \sum_{j=2}^{N}c_{j}\sqrt{q_{j}}e^{i\theta_{j}}.
\end{eqnarray}
We note that $p$ is the probability that we would determine that the state
$\psi$ is in $\mathcal{H}_{2}$, and $q$ is the probability that we 
would fail to do so.  Taking the inner product of the above equation
with $U|\psi_{1}\rangle$ gives us that
\begin{equation}
q_{1}q=|\langle\psi_{1}|\psi\rangle |^{2} .
\end{equation}
Therefore, the POVM can be used to solve a more general problem than the one
for which it was designed.  In particular, suppose we are given a system
that is guaranteed to be either in the state $\psi_{1}$ or in a state in the 
subspace $\mathcal{H}_{2}$, and we want to determine which is the case.
We can do this by applying the POVM presented here, and our probability of
failing is given by the above equation.  Because the POVM was designed for
a particular basis of $\mathcal{H}_{2}$, it will not necessarily be
optimal for the more general problem, but it will work.

\section{Application to a basis for Boolean functions}
We can now apply this result to
distinguishing between sets of Boolean functions.  Let $f(x)$,
where $0\leq x\leq 2^{n}-1$, be a Boolean function, i.e.\ $f(x)$
is either $0$ or $1$.  One of the sets we want to consider is the
set of balanced functions.  The other will consist of two ``biased''
functions.  We shall call a function biased if it is not balanced
or constant, i.e.\ if it returns $0$ on $m_{0}$ of its arguments,
$1$ on $m_{1}=2^{n}-m_{0}$, and $m_{0}\neq m_{1}\neq 0$ or $2^{n-1}$.
In order to discriminate a particular biased function from an unknown 
balanced one $2^{(n-1)}+m_{1}+1$ function evaluations are necessary in the 
worst case classically, where we have assumed, without loss of 
generality, that $m_{1}<m_{0}$.  This number comes from the fact that there
are balanced functions that agree with our biased function in $2^{(n-1)}
+m_{1}$  places.  Suppose that a balanced function is $1$ for every argument 
for which our biased function is 1.  This means that they agree for
at least $m_{1}$ arguments.  Of the remaining arguements, the biased
function will be $0$ on all of them, and the balanced function will
be $0$ on $2^{(n-1)}$ of them.  This means that the functions agree in
a total of $2^{(n-1)}+m_{1}$ places.
  
As has been mentioned,  the second set of functions we shall consider has
only two members, and we shall call it $W_{k}$. A function is in
$W_{k}$ if $f(x)=0$ for $0\leq x< [(2^{k}-1)/2^{k}]2^{n}$ and
$f(x)=1$ for $[(2^{k}-1)/2^{k}]2^{n}\leq x\leq 2^{n}-1$, or if
$f(x)=1$ for $0\leq x< [(2^{k}-1)/2^{k}]2^{n}$ and $f(x)=0$ for
$[(2^{k}-1)/2^{k}] 2^{n}\leq x\leq 2^{n}-1$. The problem we wish
to consider is distinguishing between balanced functions
and functions in $W_{k}$, that is, we are given an unknown
function that is in one of the two sets, and we want to find out
which set it is in. We note that the two functions in $W_{k}$ are
biased functions, so that this problem is a particular instance of
a more general problem of distinguishing a particular set of
biased functions from balanced functions.  This is by no means the
only example the method we are proposing here can handle, but it
is a particularly simple one.

This is clearly a variant of the Deutsch-Jozsa problem \cite{deutsch}.
In that case one is given an unknown function that is either
balanced or constant, and one wants to determine which.
Classically, in the worst case one would have to evaluate the
function $D/2+1$ times, where we have set $D=2^{n}$, but in the
quantum case only one function evaluation is necessary.  The
solution of this problem makes use of the unitary mapping
\begin{equation}
|x\rangle |y\rangle\rightarrow |x\rangle |y+f(x)\rangle ,
\end{equation}
where the first state, $|x\rangle$, is an $n$-qubit state, the
second state, $|y\rangle$, is a single qubit state, and the
addition is modulo $2$.  The state $|x\rangle$, where $x$ is an
$n$-digit binary number, is a member of the computational basis
for $n$ qubits, and the state $|y\rangle$, where $y$ is either $0$
or $1$, is a member of the computational basis for a single qubit.
In solving the Deutsch-Jozsa problem, this mapping is employed in
the following way
\begin{equation}
\label{map2} \frac{1}{\sqrt{2D}}\sum_{x=0}^{D-1}|x\rangle
(|0\rangle -|1\rangle ) \rightarrow
\frac{1}{\sqrt{2D}}\sum_{x=0}^{D-1}(-1)^{f(x)}|x\rangle (|0\rangle
-|1\rangle ) ,
\end{equation}
and we shall do the same.  This has the effect of mapping Boolean
functions to vectors in the $D$-dimensional Hilbert space, ${\cal
H}_{D}$; the final qubit is not entangled with the remaining $n$
qubits and can be discarded.  The vectors
$\sum_{x=0}^{D-1}(-1)^{f(x)} |x\rangle$ that are produced by
balanced functions are orthogonal to those produced by constant
functions.  This is why the Deutsch-Jozsa problem is easy to solve
quantum mechanically.  In our case, the vectors produced by
functions in $W_{k}$ are not orthogonal to those produced by
balanced functions.  However, we can solve the problem
probabilistically by using our unambiguous state filtering
procedure.

In order to apply quantum state filtering to this problem, we note
that both functions in $W_{k}$ are mapped, up to an overall sign,
to the same vector in ${\cal H}_{D}$, which we shall call
$|w_{k}\rangle$. The vectors that correspond to balanced functions
are contained in the subspace, ${\cal H}_{b}$, of ${\cal H}_{D}$,
where ${\cal H}_{b}=\{ |v\rangle\in {\cal H}_{D}|\sum_{x=0}^{D-1}
v_{x}=0\}$, and $v_{x}=\langle x|v\rangle$. This subspace has
dimension $2^{n}-1$, and it is possible to choose an orthonormal
basis, $\{|v_{j}\rangle |j=2,\ldots D\}$, for it in which each basis 
element corresponds to a particular balanced Boolean function. 
As any state in ${\cal H}_{b}$ is a linear combination of
the  $|v_{j}\rangle$ basis vectors, a state filtering procedure that
can optimally distinguish $|w_{k}\rangle$ from the set $\{
|v_{j}\rangle \}$ will also discriminate $|w_{k}\rangle$ from any set
of states in ${\cal H}_{b}$, though not necessarily optimally. 

Let us now construct the basis for $\mathcal{H}_{D}$ that we wish to use.
This will be done inductively.  For $D=2$, choose the basis vectors to
be
\begin{equation}
|v_{1}^{(1)}\rangle = \frac{1}{\sqrt{2}}(1,1) \hspace{1cm} 
|v_{2}^{(1)}\rangle = \frac{1}{\sqrt{2}}(1,-1) .
\end{equation}
Note that $|v_{2}^{(1)}\rangle$ is the vector that corresponds to the two
balanced one-bit Boolean functions, and $|v_{1}^{(1)}\rangle$ corresponds to the two constant functions.  These vectors are orthonormal.  For $D=2^{2}$ the basis vectors are
\begin{eqnarray}
|v_{1}^{(2)}\rangle = \frac{1}{2}(1,1,1,1) & |v_{2}^{(2)}\rangle = 
\frac{1}{2}(1,1,-1,-1) \nonumber \\
|v_{3}^{(2)}\rangle = \frac{1}{2}(1,-1,1,-1) & |v_{4}^{(2)}\rangle = 
\frac{1}{2}(1,-1,-1,1) .
\end{eqnarray}
The vector $|v_{1}^{(2)}\rangle$ corresponds to the two constant
functions, and the other basis elements correspond to balanced functions.

Now let us construct a basis for arbitrary $D=2^n$.  First, suppose that we
have all of the bases $\{ v_{j}^{m} |j=1,\ldots 2^{m}\}$ for $m\leq n-1$.
We now want to construct the basis for $D=2^n$.  Denote a string of $p$
ones by $u(p)$.  Our first basis vector is just
\begin{equation}
\label{v1}
v_{1}^{(n)}=\frac{1}{\sqrt{D}}(u(D)),
\end{equation}
which just corresponds to a constant function on $n$ bits, and our
second is
\begin{equation}
\label{v2}
v_{2}^{(n)}=\frac{1}{\sqrt{D}}(u(D/2),-u(D/2)),
\end{equation}
The general basis vector is composed of blocks of ones and minus ones.
It can be expressed as
\begin{eqnarray}
\label{vector}
\frac{1}{2^{(n-p+1)/2}}(\:
(v_{j}^{(p-1)})_{1}(u(D/2^{p}),-u(D/2^{p})), \nonumber \\
\ldots (v_{j}^{(p-1)})_{2^{(p-1)}}(u(D/2^{p}),-u(D/2^{p}))\: )\ ,
\end{eqnarray}
where $(v_{j}^{(k-1)})_{q}$ is just the $q$th component of 
$v_{j}^{(k-1)}$.  Here we have that $j$ runs from $1$ to $2^{p-1}$, and
$p$ goes from $2$ to $n-1$. That these vectors are orthogonal for
different values of $p$ can be seen rather easily.  Let the first
vector have $p=p_{1}$ and the second have $p=p_{2}$, where $p_{1}<
p_{2}$. When taking the inner product, each block of the form 
$(u(D/2^{p_{1}}),-u(D/2^{p_{1}})$, in the first vector
will be paired with a block of uniform ones or minus ones in the second
vector.  This means that each $(u(D/2^{p_{1}}),-u(D/2^{p_{1}})$  block
in the first vector will contribute a zero to the total inner product,
so that the entire inner product is itself just zero.  When taking the
inner product of two vectors for which the values of $p$ are the same, 
we see that the inner product is just proportional to 
$\langle v_{j_{1}}^{(p-1)}|v_{j_{2}}^{(p-1)}\rangle$, where $j_{1}$
and $j_{2}$ are the $j$ values for the two vectors.  Therefore, if
$j_{1}\neq j_{2}$ the vectors are orthogonal.  Henceforth, we shall
denote the vector in Eq.\ (\ref{vector}) by $v_{p,j}$ where $j$ runs 
from $1$ to $2^{p-1}$, and $p$ goes from $2$ to $n-1$.  The vector
in Eq.\ (\ref{v1}) will be denoted as $v_{0,1}$, and the vector
in Eq.\ (\ref{v2}) as $v_{1,1}$. 

Let us first see how the filtering procedure performs when
applied to the problem of distinguishing
$|w_{k}\rangle$ from the $D-1$ orthonormal basis
states, $|v_{p,j}\rangle$, for $p>0$, in  ${\cal H}_{b}$. We assume 
that their \emph{a priori} probabilities are equal, and we shall denote
this probability by $\eta$, where  $\eta = (1-\eta_{1})/(2^{n}-1)$ and
$\eta_1$ is the \emph{a priori} probability for the function to be
in $W_k$. Together with the state $|w_k \rangle$, the total number
of states is $D$, the dimension of the system Hilbert space. 

In order to find which of the three measurement procedures is optimal,
we need to calculate both $S$ and $\|w_{k}^{\parallel}\|$.  Now 
$|w_{k}\rangle$ is a unit vector, so that the sum of the square of its
component along $|v_{0,1}\rangle$ and $\|w_{k}^{\parallel}\|^{2}$ is
one.  Therefore, we have that
\begin{eqnarray}
\|w_{k}^{\parallel}\|^{2}& = &1-|\langle v_{0,1}|w_{k}\rangle |^{2}
\nonumber \\
 & = & \frac{2^{k}-1}{2^{(2k-2)}} .
\end{eqnarray}
The calculation of $S$ is particularly simple in this case.  We have
\begin{eqnarray}
S & = & \eta\sum_{p=1}^{n-1}\sum_{j=1}^{2^{p-1}}|\langle v_{p,j}|w_{k}
\rangle |^{2} \nonumber \\
 & = & \eta (1-|\langle v_{0,1}|w_{k}\rangle |^{2}) = \frac{\eta (2^{k}-1)}
{2^{(2k-2)}} \nonumber \\
{}& = & \eta \|w_{k}^{\parallel}\|^{2} \ .
\end{eqnarray}
Substituting these quantities into the conditions in Eq.\ (\ref{Qmin}),
we find that the first von Neumann measurement is optimal if $\eta_{1} >
\zeta_{1}$, the POVM is optimal if $\zeta_{2}\leq \eta_{1} \leq \zeta_{1}$,
and the second von Neumann measurement is optimal if $\eta_{1}<\zeta_{2}$
where
\begin{eqnarray}
\zeta_{1} & = & \left[ 1+\frac{(2^{n}-1)(2^{k}-1)}{2^{2(k-1)}}\right]^{-1}
\simeq 2^{-(n-k+2)} \nonumber \\
\zeta_{2} & = & \frac{2^{k}-1}{2^{2(k-1)}(2^{n}-1)+2^{k}-1} \simeq 
2^{-(n+k-2)} ,
\end{eqnarray}
where, in the approximate expressions, we have assumed that $2^{n}\gg
2^{k}\gg 1$.  

It is useful to get an idea of how much the failure probabilities of the 
different prodecures differ.  To do so, we will look in the range in which 
the POVM is optimal, and compare the failure probabilities of the
three different measurements.  In this range, the failure probabilities
are given by
\begin{eqnarray}
\label{errors}
Q_{POVM}& = & \frac{1}{2^{k-2}}\left[\frac{\eta_{1}(1-\eta_{1})
    (2^{k}-1)} {2^{n}-1}\right]^{1/2} , \nonumber \\
Q_{1} & = & \eta_{1}+\frac{(1-\eta_{1})(2^{k}-1)} {2^{2k-2}
    (2^{n}-1)} , \nonumber \\
Q_{2} & = & \frac{\eta_{1}(2^{k}-1)}{2^{2k-2}} +
    \frac{1-\eta_{1}} {2^{n}-1} .
\end{eqnarray}
It should be noted that while $Q_{POVM}$ in the above expression is
valid only if $\zeta_{2}\leq \eta_{1} \leq \zeta_{1}$, the expressions
for $Q_{1}$ and $Q_{2}$ are valid for any value of $\eta_{1}$.
Assuming that $2^{n}\gg 2^{k}\gg 1$ we find that
\begin{equation}
\label{ratio}
\frac{Q_{POVM}}{Q_{1}} \simeq \frac{4}
     {\sqrt{2^{n+k}\eta_{1}}} \hspace{1cm} 
\frac{Q_{POVM}}{Q_{2}} \simeq 4\sqrt{2^{n-k}\eta_{1}} ,
\end{equation}
so that the POVM result represents a considerable improvement over
either of the von Neumann measurements when it is optimal.  For example, 
in the case
in which all of the \emph{a priori} probabilities are equal, i.e.\
$\eta_{1}=1/2^{n}$, we have that
\begin{equation}
Q_{POVM}=\frac{(2^{k}-1)^{1/2}}{2^{n+k-2}}\simeq \frac{1}{2^{n-2+(k/2)}} ,
\end{equation}
both of the ratios in Eq.\ (\ref{ratio}) are $4/2^{k/2}$.
For $ k\gg 1$, this implies that the difference in performance
between the POVM and the von Neumann measurements can be
significant.

\section{Example of POVM}
So far, our discussion of the POVM has been rather abstract; we know
when it exists and when it does not, and we have described how it 
works.  It is useful, however, to see explicitly how the unitary
operator (see Eq.\ (\ref{unit})) that plays a prominent role in the 
scheme can be constructed.  Our task is to find explicit expressions
for the vectors $|\psi_{j}^{\prime}\rangle$.  In order to do this, we 
shall consider an explicit example, the case $k=2$.

We first note that $\langle w_{2}|v_{p,j}\rangle = 0$ for all $p>2$, 
which means that all of these vectors can be perfectly distinguished 
from $|w_{2}\rangle$.  Therefore, we define $U|v_{p,j}\rangle =
|v_{p,j}\rangle$ for $p>2$.  $U$ will map the remaining
four-dimensional subspace, spanned by the vectors $|v_{p,j}\rangle$, 
with $p\leq 2$, into itself, and we can henceforth confine our attention 
to this subspace.

Let us now set
\begin{eqnarray}
|\psi_{1}\rangle = |w_{2}\rangle & |\psi_{2}\rangle = |v_{1,1}\rangle
\nonumber \\
|\psi_{3}\rangle = |v_{2,1}\rangle & |\psi\rangle = |v_{2,2}\rangle 
\end{eqnarray}
The condition that guarantees the existence of $U$ is that the matrix,
$M$, given by
\begin{equation}
M_{jk}=\langle\psi_{j}^{\prime}|\psi_{k}^{\prime}\rangle =\langle\psi_{j}|
\psi_{k}\rangle - \sqrt{q_{j}q_{k}}e^{i(\theta_{k}-\theta_{j})} ,
\end{equation}
be positive.  Taking the inner product of $U|\psi_{j}\rangle$ and
$U|\psi_{1}\rangle$, $j\neq 1$, we find that
\begin{equation}
\langle\psi_{j}|\psi_{1}\rangle = \sqrt{q_{1}q_{j}}e^{i(\theta_{1}
-\theta_{j})} ,
\end{equation}
so that we can express $M_{jk}$, for $j,k>1$, as
\begin{equation}
M_{jk}=\langle\psi_{j}|\psi_{k}\rangle -\frac{1}{q_{1}}\langle\psi_{j}|
\psi_{1}\rangle\langle\psi_{1}|\psi_{k}\rangle .
\end{equation}
Setting $x=1/(4q_{1})$, we find that
\begin{equation}
M=\left(\begin{array}{cccc} 1-q_{1} & 0 & 0 & 0 \\ 0 & 1-x & -x & x \\
0 & -x & 1-x & x \\ 0 & x & x & 1-x \end{array}\right) .
\end{equation}
The eigenvalues of this matrix are $1-q_{1}$, $1-[3/(4q_{1})]$, and $1$,
which is doubly degenerate.  The matrix is positive if $(3/4)\leq
q_{1}\leq 1$.

We can find the vectors $|\psi_{j}^{\prime}\rangle$ by noting that,
because $M$ is positive, it can be expressed as $M=B^{\dagger}B$,
where $B=U_{0}\sqrt{M}$, and $U_{0}$ is an arbitrary unitary operator.
If we set
\begin{equation}
|\psi_{j}^{\prime}\rangle = B_{1j}|v_{0,1}\rangle+B_{2j}|v_{1,1}\rangle
+B_{3j}|v_{2,1}\rangle + B_{4j}|v_{2,2}\rangle ,
\end{equation}
then we find that $M_{jk}=\langle\psi_{j}^{\prime}|\psi_{k}^{\prime}
\rangle$, and we have completely specified $U$.  Note that there is
considerable arbitrariness, because we are free to choose the unitary
operator $U_{0}$.  A particular choice yields
\begin{eqnarray}
|\psi_{1}^{\prime}\rangle & = & \sqrt{1-q_{1}}|v_{0,1}\rangle ,
\nonumber \\
|\psi_{2}^{\prime}\rangle & = & \frac{1}{\sqrt{2}}|v_{1,1}\rangle
+\frac{1}{\sqrt{6}}|v_{2,1}\rangle + \sqrt{1-\frac{x}{3}}|v_{2,2}\rangle
\nonumber \\
|\psi_{3}^{\prime}\rangle & = & -\frac{1}{\sqrt{2}}|v_{1,1}\rangle
+\frac{1}{\sqrt{6}}|v_{2,1}\rangle + \sqrt{1-\frac{x}{3}}|v_{2,2}\rangle
\nonumber \\
|\psi_{4}^{\prime}\rangle & = & \sqrt{\frac{2}{3}}|v_{2,1}\rangle - 
\sqrt{1-\frac{x}{3}}|v_{2,2}\rangle .
\end{eqnarray}
These vectors, plus the choice of $q_{1}$, which one finds by minimizing
the failure probability and checking to see whether it is in the allowed
range (in this case between 3/4 and 1), completely specify the POVM.

\section{Application to all balanced functions}
Now that we know how this procedure performs on the basis vectors
in ${\cal H}_{b}$, we shall examine its performance on any even 
function.  What we shall do is compare the performance of the two
von Neumann measurements to that of the POVM that is optimal for
the basis vectors.  We cannot apply the filtering procedure directly
to the set of balanced functions, unless we want the dimension of the
space in which the POVM is defined to be equal to the number of
balanced functions.  We are instead interested in a POVM acting
in the space $\mathcal{H}_{D}\oplus\mathcal{H}_{A}$, where 
$\mathcal{H}_{A}$ is a one-dimensional auxiliary Hilbert space.  This
can be accomplished by using the POVM that was derived for the more
restricted problem of distinguishing the basis vectors from 
$|w_{k}\rangle$.

First let us see how the von Neumann measurements perform.  We shall assume
that $|w_{k}\rangle$ has an \emph{a priori} probability of $\eta_{1}$
and that each of the vectors corresponding to a balanced function has
the same probability.  There are
\begin{equation}
M_{bal}=\left(\begin{array}{c} D \\ D/2 \end{array}\right) ,
\end{equation}
vectors corresponding to balanced functions, so that the probability of
each of them is $(1-\eta_{1})/M_{bal}$.  In order to calculate the
average error probabilities we need to calculate
\begin{equation}
\label{sum}
S_{b}=\frac{1-\eta_{1}}{M_{bal}}\sum_{v_{b}}|\langle w_{k}|
v_{b}\rangle |^{2} ,
\end{equation}
where the sum is over all vectors corresponding to balanced functions.
This sum is calculated in the Appendix, and we find that
\begin{equation}
\label{sb}
S_{b}=\frac{1-\eta_{1}}{D-1}f_{k} ,
\end{equation}
where $f_{k} = (2^{k}-1)/2^{2k-2}$. This expression is identical to 
the one calculated for the basis vectors alone.  That means that the
failure probabilities for the two von Neumann measurements, in the
case of all balanced functions and not just the basis vectors, are 
still given by the expressions in Eq.\ (\ref{errors}).

Now let us turn our attention to the POVM.  As discussed at the end of
Section 2, even for the POVM designed to distinguish the basis vectors
from $|w_{k}\rangle$, the failure probability for any balanced function
vector, $|v_{b}\rangle$, is given by
\begin{equation}
q_{v_{b}}=\frac{|\langle w_{k}|v_{b}\rangle |^{2}}{q_{1}} ,
\end{equation}
so that the total failure probability is
\begin{equation}
Q_{POVM}=\eta_{1}q_{1}+\frac{S_{b}}{q_{1}} .
\end{equation}
Choosing $q_{1}$ to minimize the right-hand side, we obtain the 
expression for $Q_{POVM}$ given in Eq.\ (\ref{errors}).  Again, we
must have $\zeta_{2}\leq \eta_{1} \leq \zeta_{1}$ for this expression
to be valid.

Now let us look at some choices for $\eta_{1}$.  If all of the functions
are equally probable, then $\eta_{1}=1/(M_{bal}+1)$, and the first von
Neumann measurement is the optimal one.  The measurement fails if the
vector is $|w_{k}\rangle$, but this event is so unlikely that it has 
a neglgible effect on the average failure probability.  If $\eta_{1}
=1/D$, that is the probabilities of getting $|w_{k}\rangle$ or
a balanced-function vector are proportional to the dimensions of the 
subspaces in which they lie, then the POVM is optimal.  If $\eta_{1}
=1/2$, so that we are equally likely to be given $|w_{k}\rangle$ or
a balanced-function vector, then the second von Neumann measurement is
optimal.  This illustrates how the best strategy is influenced by the
\emph{a priori} probabilities.

\section{Conclusion}
If one is given a Boolean function that is promised to be either even or
in $W_{k}$, classically, in the worst case, one would have to evaluate it
$2^{n}[(1/2)+(1/2^{k})]+1$ times to determine to which set it belongs. 
Using quantum information processing methods, one has a very good
chance of determining this with only one function evaluation. This
shows that Deutsch-Jozsa-type algorithms need not be limited to
constant functions; certain kinds of biased functions can be
discriminated as well.

Unambiguous state discrimination is a procedure that is of
fundamental interest in quantum information theory.  Its only
application so far has been to quantum cryptography.  The results
presented here suggest that related methods can also serve as a
tool in the development of quantum algorithms.

\section*{Acknowledgments}
We would like to thank Ulrike Herzog for useful conversations.
This research was supported by the Office of Naval Research (Grant
No. N00014-92-J-1233), the National Science Foundation (Grant
No. PHY-0139692), the Hungarian Science Research Fund (Grant No. T
03061), a PSC-CUNY grant, and a CUNY collaborative grant.

\section*{Appendix}
We want to evaluate the sum in Eq.\ (\ref{sum}).  The vector $|w_{k}
\rangle$ is given by
\begin{equation}
|w_{k}\rangle = \frac{1}{\sqrt{D}}(1,1,\ldots 1, -1,-1,\ldots -1) ,
\end{equation}
where the first $D-(D/2^{k})$ places are $1$'s and the final $D/
2^{k}$ places are $-1$'s.  Now consider a vector corresponding to
a balanced function, which has $m$ $1$'s in its last $D/2^{k}$ places,
where $0\leq m \leq D/2^{k}$, so that it has $(D/2)-m$ $1$'s in its
first $D-(D/2^{k})$ places.  The overlap of $|w_{k}\rangle$ with this 
vector is
\begin{equation}
\frac{1}{2^{k-1}}-\frac{4m}{D} .
\end{equation}
The number of balanced functions of this type, $C_{m}$, is given by
\begin{equation}
C_{m}=\left(\begin{array}{c}D/2^{k} \\ m \end{array}\right)
\left(\begin{array}{c}D(2^{k}-1)/2^{k} \\ (D/2)-m \end{array}\right) ,
\end{equation}
so that the sum we have to evaluate is given by
\begin{equation}
\label{sbsum}
S_{b}=\eta\sum_{m=0}^{D/2^{k}}C_{m}\left( \frac{1}{2^{k-1}}
-\frac{4m}{D}\right)^{2} ,
\end{equation}
where $\eta =(1-\eta_{1})/M_{bal}$ is the \emph{a priori} probability
of each balanced function.

In order to find $S_{b}$, we have to evaluate three types of sums.  We
shall discuss one is some detail, and simply give results for the other
two.  The first sum is
\begin{equation}
s_{0}=\sum_{m=0}^{D/2^{k}}C_{m}
\end{equation}
This can be evaluated by noting that
\begin{eqnarray}
(1+x)^{D/2^{k}}=\sum_{m=0}^{D/2^{k}}\left(\begin{array}{c}D/2^{k} \\ m 
\end{array}\right) x^{m} \nonumber \\
(1+x)^{D(2^{k}-1)/2^{k}}=\sum_{l=0}^{D(2^{k}-1)/2^{k}}
  \left(\begin{array}{c} 
D(2^{k}-1)/2^{k} \\ (D/2)-m \end{array}\right) x^{l} ,
\end{eqnarray}
Multiplying these two expressions together we find 
\begin{eqnarray}
(1+x)^{D}{}&=&\sum_{m=0}^{D/2^{k}}\sum_{l=0}^{D(1-2^{-k})}
\left(\begin{array}{c}D/2^{k} \\ m \end{array}\right) \nonumber \\ 
{}&&\left(\begin{array}{c}D(1-2^{-k}) \\ (D/2)-m \end{array}\right) 
x^{l+m} \ .
\end{eqnarray}
Comparing coefficients of $x^{D/2}$ on both sides of this equation,
we find that
\begin{equation}
s_{0}=\left(\begin{array}{c}D \\ D/2 \end{array}\right) .
\end{equation}
The remaining two sums are
\begin{equation}
s_{1}=\sum_{m=0}^{D/2^{k}}mC_{m}=\frac{D}{2^k}\left(\begin{array}{c}
D-1 \\ (D/2)-1 \end{array}\right) ,
\end{equation}
and
\begin{eqnarray}
s_{2}&=&\sum_{m=0}^{D/2^{k}}m^{2}C_{m} \nonumber \\
{}&=&
\frac{D}{2^k}\left(\frac{D}{2^k} 
-1\right)\left(\begin{array}{c}D-2 \\ (D/2)-2 \end{array}\right)
\nonumber \\
{}&&+\frac{D}{2^k}\left(\begin{array}{c}D-1 \\ (D/2)-1
\end{array}\right). 
\end{eqnarray}
Substitution of these expressions into Eq.\ (\ref{sbsum}) yields the 
result in Eq.\ (\ref{sb}).

\bibliographystyle{unsrt}

\end{document}